\documentclass[preprint,12pt]{elsarticle}

\usepackage{amssymb, amsmath, amsfonts}
\usepackage{ulem}
\usepackage{amsthm}

\usepackage{graphicx}
\usepackage{textcomp}
\usepackage{array}
\newcolumntype{P}[1]{>{\centering\arraybackslash}p{#1}}
\newcolumntype{M}[1]{>{\centering\arraybackslash}m{#1}}
\usepackage{color}
\usepackage{multirow}
\usepackage{makecell}
\usepackage{tabularx}
\usepackage{hyperref}

\journal{Photoacoustics}
\begin{document}
\begin{frontmatter}

\title{Learning-based sound speed estimation and aberration correction for linear-array photoacoustic imaging}

\author[1]{Mengjie Shi}
\author[1]{Tom Vercauteren}
\author[1]{Wenfeng Xia\corref{cor1}}
\ead{wenfeng.xia@kcl.ac.uk}
\cortext[cor1]{Corresponding author}
\affiliation[1]{organization={School of Biomedical Engineering and Imaging Sciences},
            addressline={King's College London}, 
            city={London},
            postcode={SE1 7EH}, 
            country={United Kingdom}}

\begin{abstract}
Photoacoustic (PA) image reconstruction involves acoustic inversion that necessitates the specification of the speed of sound (SoS) within the medium of propagation. Due to the lack of information on the spatial distribution of the SoS within heterogeneous soft tissue, a homogeneous SoS distribution (such as 1540 m/s) is typically assumed in PA image reconstruction, similar to that of ultrasound (US) imaging. Failure to compensate the SoS variations leads to aberration artefacts, deteriorating the image quality. Various methods have been proposed to address this issue, but they usually involve complex hardware and/or time-consuming algorithms, hindering clinical translation. In this work, we introduce a deep learning framework for SoS estimation and subsequent aberration correction in a dual-modal PA/US imaging system exploiting a clinical US probe. As the acquired PA and US images were inherently co-registered, the estimated SoS distribution from US channel data using a deep neural network was incorporated for accurate PA image reconstruction. The framework comprised an initial pre-training stage based on digital phantoms, which was further enhanced through transfer learning using physical phantom data and associated SoS maps obtained from measurements. This framework achieved a root mean square error of 10.2 m/s and 15.2 m/s for SoS estimation on digital and physical phantoms, respectively and structural similarity index measures of up to 0.86 for PA reconstructions as compared to the conventional approach of 0.69. A maximum of 1.2 times improvement in signal-to-noise ratio of PA images was further demonstrated with a human volunteer study. Our results show that the proposed framework could be valuable in various clinical and preclinical applications to enhance PA image reconstruction. 
\end{abstract}

\end{frontmatter}

\section{Introduction}
PA imaging is a hybrid modality that combines rich optical contrast from optical imaging, with high spatial resolution and large imaging depths from US imaging. In the past two decades, PA imaging has demonstrated great potentials for a wide range of applications in preclinical and clinical settings \cite{beard2011biomedical,zhao2019minimally,wang2016practicala}. PA involves the illumination of biological tissues from pulsed or modulated continuous-wave light sources such as solid-state lasers or light-emitting diodes (LEDs). The light is then locally absorbed by endogenous chromophores such as hemoglobin, water and lipids, and exogenous contrast agents. This absorption leads to local temperature increases and transient thermal expansion, serving as a source of US wave generation. The amplitude, frequency, and time-of-flight of the US signal, respectively, provide information of the optical absorption, size and spatial location of the optical absorbers. In PA tomography configurations, the generated US waves propagate through tissue and are then detected with US sensors located on the tissue surface. Subsequently, acoustic inversion is performed on the acquired time series US data to reconstruct PA images representing optical absorption contrast. For this purpose, the SoS in the propagation medium is usually assumed to be homogeneously distributed (typically 1540 m/s as an average for soft tissue). However, SoS of soft tissue is highly dependent on tissue types, varying from around 1450 m/s (fat) to 1580 m/s (muscle) \cite{mamou2013quantitative}. Besides, SoS inconsistency can exist between the acoustic coupling medium and the tissue. As such, this assumption can lead to significant US aberration artefacts, degrading the image contrast and spatial resolution \cite{dean-ben2014effectsa}. 

Various methods were proposed to mitigate the SoS-induced aberration artefacts in PA tomography \cite{treeby2011automatic,yoon2012enhancement,anastasio2005halftime,jiang2006spatially,zhang2006reconstructiona,zhang2008simultaneousa,huang2016joint,huang2013fullwave,matthews2018parameterized,matthews2017joint,jeon2021deep,chen2013tr,cai2019feature,deng2022multisegmenteda}. In 2011, Treeby et al. reported an autofocus algorithm to arrive at an optimal single SoS value for the medium by iterative optimisation \cite{treeby2011automatic}. The automated selection of the optimal SoS can be achieved by maximising an image resolution metric such as image sharpness \cite{treeby2011automatic} and a coherent factor based on delay-compensated US channel data in time domain \cite{yoon2012enhancement}. These methods implemented a global optimisation of SoS. Image resolution metrics were thus optimised over the whole focus area and local variations on SoS within heterogeneous tissue were neglected. To incorporate the heterogeneity effect, parameterised SoS maps can be directly reconstructed using PA measurements. Several works investigated concurrent recovery of both SoS distributions and initial pressure distributions or optical absorption from PA measurements, which was referred to as a joint reconstruction (JR) problem \cite{jiang2006spatially,zhang2006reconstructiona, zhang2008simultaneousa}. Jiang et al. reported a JR approach by seeking numerical solutions to the Helmholtz equation using a finite element method \cite{jiang2006spatially}. Zhang et al. proposed a time-domain method based on explicitly exploring two-fold data redundancy in a generalized Radon transform imaging model \cite{zhang2006reconstructiona}. Zhang et al. further reported an implicit method by iteratively optimising a cost function with respect to the SoS map and optical absorption simultaneously \cite{zhang2008simultaneousa}. In fact, accurate JR may not be achievable due to the numerical instability, as shown in \cite{huang2016joint} and \cite{stefanov2013instability}. Therefore, prior information of SoS distributions and detection geometries was usually incorporated for more accurate JR \cite{matthews2018parameterized}\cite{kirsch2012simultaneous}. Cai et al. proposed a feature coupling (FC) based JR method integrated with a full ring array photoacoustic computational tomography (PACT) system \cite{cai2019feature}. The SoS distributions were obtained through an iterative process that maximised the similarity between two PA images reconstructed by the two half-ring data. Validation on an in vivo mouse liver model demonstrated its superiority on distortion mitigation. However, the performance of the FC method can be significantly deteriorated when the image sparsity is high. To address this issue, a multi-segment feature coupling (MSFC) method was developed. The full ring array was partitioned into finer groups, and the FC method was applied to each group individually to estimate a direction-specific SoS \cite{deng2022multisegmenteda}. 

SoS distributions can also be independently measured with US transmission tomography and used for correcting PA image reconstruction in dedicated systems combining US and PA measurements \cite{xiaAcousticspeedCorrectionPhotoacoustic2014}. In adjunction to PACT, US tomography can be implemented in a transmission mode by positioning a US transmitter and a receiver in an opposite position \cite{zhang2006reconstructiona}. Passive elements can be used as US transmitters in a US transmission tomography configuration based on the PA effect in a PA imaging system by sharing the same US receiver \cite{manohar2007concomitant, jose2011passive}. Mercep et al. proposed a transmission-reflection optoacoustic US (TROPUS) imaging platform that can retrieve multiple acoustic properties including SoS, acoustic attenuation and reflectivity \cite{mercepTransmissionReflectionOptoacoustic2019}. In such hybrid systems, SoS maps acquired via US transmission tomography are employed for optimising PA image quality during the reconstruction. However, these methods are usually associated with a high degree of complexity, both in the hardware design and algorithm developments. 

Pulse-echo US can be readily integrated with PA imaging by sharing a clinical US array probe for real-time imaging. Such dual-modal imaging systems can provide complementary morphological and molecular information of tissue based on optical absorption. Therefore, this configuration could facilitate clinical adoption of PA imaging and has thus attracted great attention \cite{kuniyilajithsingh2016handheld, shi2022enhanced,shi2022improving,xia2018handheld}. With recent advances in deep learning (DL), Jeon et al. proposed a DL-based framework for mitigating SoS aberration and streak artefacts resulting from sparse sampling in a linear-array PA imaging system \cite{jeon2021deep}. However, an explicit SoS distribution was not accessible for direct evaluation. 

In this work, we propose a learning-based method that retrieves SoS distributions using US radio-frequency (RF) channel data to inform PA image reconstruction in a dual-modal PA/US imaging system based on a clinical probe. To achieve this, we trained a deep neural network to estimate SoS distributions from US RF channel data obtained from digital phantoms with 2D simulations. Further, we improved its performance by conducting transfer learning using data from physical phantoms. The performance of the framework including SoS estimation and aberration correction in PA imaging was evaluated on digital phantoms with previously unseen structures and echogenicities, agar-based tissue-mimicking phantoms, ex vivo porcine tissues, and in vivo human vasculature. Therefore, we demonstrated the viability of utilising SoS information inherent in co-registered US data for enhancing PA reconstruction, without time-consuming iterative algorithms or complex hardware developments, thus paving the way for clinical translation of linear-array-based PA imaging techniques.

\label{deep-neural-networks}
\begin{figure*}[!ht]
\centerline{\includegraphics[width= \textwidth]{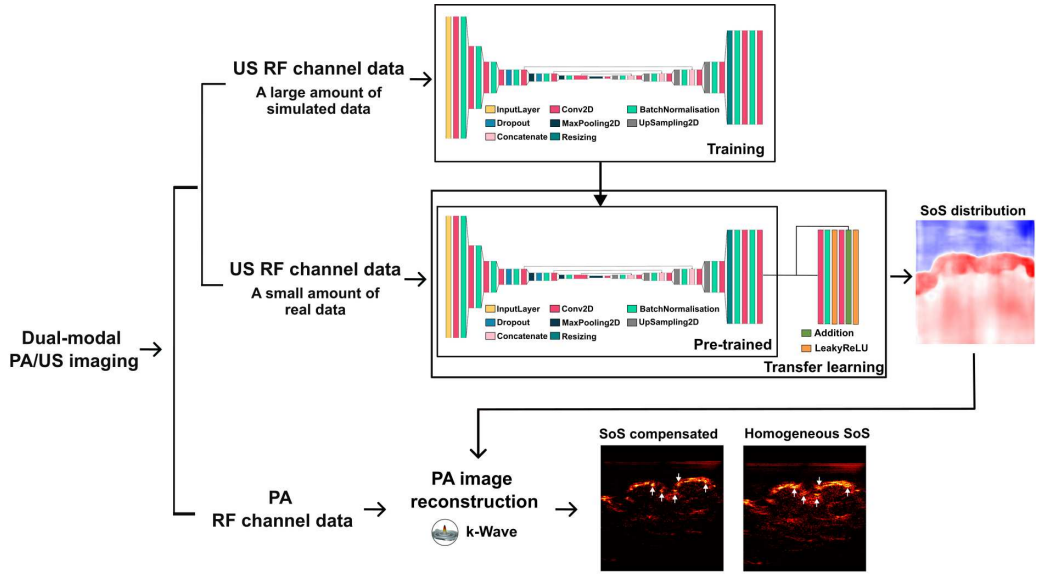}}
\caption{Learning-based sound speed estimation and aberration correction framework for dual-modal photoacoustic (PA) /ultrasound (US) imaging. Top branch: Deep learning based speed of sound (SoS) estimation using US channel data. Bottom branch: SoS compensation for PA image reconstruction (PA images were acquired from cross sections of human fingers in vivo).}
\label{fig1-workflow}
\end{figure*}

\section{Materials and Methods}
This section is structured as follow: Sec. \ref{ultrasound-simulations} details the digital simulations used for generating US data for training. Sec.\ref{deep-neural-networks} describes the establishment of the deep learning model used to estimate SoS distribution from US channel data, including a pre-training stage on US simulations and a transfer learning on small amounts of real data. Sec. \ref{pa-image-reconstruction} introduces a PA image reconstruction method that incorporates the SoS distribution obtained via deep learning. Sec.\ref{model-evaluation} details the experimental validation of the model including data acquisition and processing methods for qualitative and quantitative analysis. 

\subsection{Ultrasound simulations}
\label{ultrasound-simulations}
To circumvent the lack of a large amount of real US data with explicit SoS distributions, training data was prepared by US simulation in 2D using the k-Wave MATLAB toolbox\cite{treeby2010kwave}. A clinical linear array US probe was simulated. The probe had 128 elements spanning 38.4 mm with a pitch size of 0.3 mm and a central frequency of 7 MHz. The dimensions of the simulation grid were 1536 $\times$ 1536 with a spatial resolution of 0.025 mm. The probe was located at one side of the grid, with 11 grid points per piezo element and 1 grid point per kerf (gap between two adjacent elements). The excitation pulses for single plane wave transmissions were 2-cycle tone burst signals with a central frequency of 7 MHz. 

The US anatomies were simulated based on the tissue models of organs and lesions used in \cite{jush2020dnnbased}. Ellipses representing US heterogeneities with various dimensions and orientations were randomly distributed within a background medium. The dimensions of the medium were 38.4 mm $\times$ 38.4 mm. The long axis of the ellipses spanned from 0 mm to (38.4 $\times \sqrt{2}$) mm while the short axis ranged from 0 mm to (19.2 $\times\sqrt{2}$ ) mm. SoS values were randomly chosen from a uniform distribution $\mathcal{U}\textsubscript{[1400, 1600]}$ for the homogeneous background and from a [1\% - 7\%] higher range for the inclusions. The echogenicity was considered by simulating the elliptical-shaped inclusions as being hyperechoic, which was demonstrated to have promising generalisation performance across different echogenicity patterns \cite{jush2022deepa}. This was achieved by either increasing the SoS values or the density of the speckles inside the inclusions for an enhanced contrast \cite{agrawal2021modeling}. For the former implementation, the hyperechoic features were simulated by randomly assigning the SoS values with [7\% - 11\%] increments to the background to 10 \% of the grid points. Acoustic attenuation and mass intensity were fixed to 0.5 dB/(MHz$\cdot$cm) and 1020 kg/m\textsuperscript{3} according to the average values of human soft tissues. The speckle density had a mean distribution of 3 speckles per $\lambda$\textsuperscript{2} ($\lambda$ is the wavelength of the transmit pulse). The intensity was assigned by sampling the scatterers with a uniform distribution $\mathcal{U}\textsubscript{[-0.03, 0.03]}$. A time gain compensation of 0.5 dB/MHz/cm at 1540 m/s was implemented. The acquired multichannel RF data were normalised to have a mean of 0 and a standard deviation of 1 for each channel. 

Furthermore, thermal noise and system noise were considered with the RF data. Thermal noise resulting from electrons’ agitation in US imaging systems was modelled using a white Gaussian noise. The noise amplitude at each channel was determined by the signal-to-noise ratio (SNR) that was randomly sampled from -80 dB to -40 dB. System noise associated with transmission interference was sampled from real US measurements. The system noise was obtained by extracting the signals corresponding to the first 50 time steps of the US RF channel data. The US simulations of 6000 samples took around 2 days with an NVIDIA Quadro RTX 5000 GPU. 

\subsection{Deep neural networks for SoS estimation}
The deep learning model was tailored for the simulated US dataset from a model proposed in \cite{feigin2020deep, jush2022deepa}. The input and output were defined as: 
\begin{align}
  \Lambda: C_{n\times m}\mapsto S_{p\times q}  
\end{align}

The RF channel data $C$ acquired through US plane wave imaging using a single plane wave transmission were taken as the input with a size of $n\times m$, where $n$=128 is the number of channels and $m$=1024 is the number of time steps at a sampling rate of 20 MHz. The output is the corresponding SoS map $S$ with dimensions of 384$\times$384 ($p\times q$) and a spatial resolution of 0.1 mm. 

As depicted in Fig. \ref{fig1-workflow}, the main model was based on a fully convolutional neural network in an encoder-decoder configuration. For the encoder, strided convolutional layers were adopted at the first three layers for accommodating the non-square input size and improving the smoothness of the SoS predictions, followed by LeakyReLU, and Batch Normalisation (BN). The next four layers consisted of convolution, LeakyReLU, MaxPooling, and BN. The decoder path had four layers consisting of convolution, LeakyReLU, bilinear upsampling, and BN. After resising the output, 1$\times$1 convolution was applied to generate the final SoS map. The encoding and decoding paths were connected by combining the output feature maps of the layer 5, 6, 7 to the corresponding layer 9, 10, 11. The model was trained on 6000 samples with a train/valid spilt of 0.9. Mean Square Error (MSE) was used as the loss function. Stochastic gradient descent (SGD) with a mini batch size of 10 and a learning rate of 0.0001 was used for training. After 100 epochs, the network converged to a Root Mean Square Error (RMSE) of 22.90 m/s on the training set and 26.74 m/s on the validation set, respectively.

Real data have characteristics (e.g. US speckles) that may not be faithfully represented by digital simulations which may have negative effect on the performance of the model. Moreover, the model can be sensitive to out-of-plane artefacts and laser-induced noise, which were not incorporated in the digital simulations. Therefore, a transfer learning of the previously trained model was proposed. As illustrated in Fig. \ref{fig1-workflow},  the transfer learning was performed with a small amount of real data acquired from agar-based tissue-mimicking phantoms. These phantoms made with agar (Agar powder, Sigma-Aldrich, Germany) and glass beads (0–63 $\mu$m, Boud Minerals Limited, UK) were prepared following a standardised protocol \cite{souza2016standard}. The reference SoS value for each agar concentration was measured using an insertion method and used for generating the corresponding SoS distribution for transfer learning (see Supplementary Materials, Sec. 1). The corresponding PA/US RF channel data were obtained with a commercially available LED-based PA/US imaging system (AcousticX, CYBERDYNE INC, Tsukuba, Japan). The base model was augmented by adding a residual block on top of its output to further improve its performance with real data \cite{he2015deepa}. During the transfer learning, the networks that were pre-trained with a large amount of simulated data were frozen. The parameters of the augmented layers were updated using the MSE as the loss function and SGD. The real dataset consisting of 48 samples was used for training with a train/valid spit of 0.9. The training was stopped after 20 epochs when the MSE was not decreasing on both the train and validation set. All the experiments were conducted on NVIDIA DGX cluster equipped with 8 A100 GPUs.

\subsection{PA image reconstruction}
\label{pa-image-reconstruction}
The SoS distributions estimated by the DL model were sequentially used to inform PA image reconstruction using a time-reversal algorithm implemented in the k-Wave MATLAB toolbox\cite{treeby2010photoacoustic}. A 39.4 mm (depth) $\times$ 38.4 mm (probe face) field with a spatial resolution of 0.05 mm was constructed. The probe used for B-mode US imaging was simulated for PA imaging. The experimental PA RF channel data with a size of 128 $\times$ 1024 were upsampled through linear interpolations to a size of 640 $\times$ 4096 given the numerical stability constrained by the simulation grid. The SoS of the medium was parameterised using the SoS distribution predicted from the corresponding US data by the DL model. 

\subsection{Framework evaluation}
\label{model-evaluation}
The proposed framework was evaluated using data acquired from digital simulations, agar-based tissue-mimicking phantoms, ex vivo tissues, and healthy human volunteers, respectively. 

Anatomies and echogenicity differing from the training data were utilised for data simulation for evaluation. These included layered structures with straight and deformed boundaries and representative echogenicity. Fig. \ref{fig2-digitalphantom}(a) shows several exemplars of the simulated B-mode US images. A total of 120 samples (40 samples from each representative pattern in Fig. \ref{fig2-digitalphantom}) were generated. The thickness of each layer and its boundary shape were randomly initialised and the SoS of each layer was randomly chosen from [1400 - 1600] m/s. The echogenicities for each layer and inclusions were randomised as isoechoic, anechoic, hypoechoic, or hyperechoic. For simplicity, the point optical absorbers were only defined in PA simulations, where a series of 2-D grid coordinates were assigned a constant initial pressure. 

The model was then tested on real data acquired with agar-based tissue-mimicking phantoms, needle (20G, BD, USA) insertions into ex vivo porcine tissues simulating minimally invasive procedures, and in vivo human vasculature. The tissue-mimicking phantoms were composed of two layers of agar phantoms with the same or different concentrations of agar and glass beads (2$\%$ w/v, 4$\%$ w/v, and 6$\%$ w/v for agar; 1$\%$ w/v and 0.5$\%$ w/v for glass beads). Three pencil leads (Faber-Castell, Stein, Germany) with a diameter of 0.5 mm were randomly positioned in the phantoms as optical absorbers. With ex vivo experiments, SoS aberration artefacts around the needle tip were demonstrated with out-of-plane and in-plane insertions, respectively. For the out-of-plane insertion, the PA signals of the needle tip was enhanced at deep depths by integrating a fibre-optic US transmitter within the needle lumen \cite{shi2022enhanced}.  Furthermore, the framework was tested using in vivo 3D human vasculature by scanning the wrist of a healthy volunteer. The human volunteer experiments were approved by the King’s College London Research Ethics Committee (study reference: HR-18/19-8881). The probe scanned over a length of 4 cm on a human wrist using a linear translation stage. A total of 1536 frames were obtained. Frame averaging was implemented across 4 consecutive frames to improve the SNR, resulting in 384 frames for model evaluation.

The PA image reconstructions with DL-based SoS compensation were compared with the reconstruction based on an autofocus algorithm \cite{treeby2011automatic} using digital phantoms, where SoS aberration was compensated using a global SoS value. To quantify the errors of SoS estimations by the model, root mean square error (RMSE) was employed. Besides, local structure similarity index measure (SSIM) was used for evaluating the PA reconstruction quality before and after aberration correction. For phantom, ex vivo, and in vivo evaluation, estimated SoS maps and the corresponding aberration-compensated PA images reconstructed before and after transfer learning were compared. Further, PA images reconstructed using the proposed DL method were compared with those from the conventional method that assumes an uniform SoS value, e.g., 1540 m/s for soft tissues. In this work, water was used as the coupling medium for all the experiments. The temperature of the coupling layer was individually measured before and after imaging for SoS determination. The enhancement in PA reconstruction quality was quantified using SNR. For reconstructed PA images, the pixels with an intensity below 0.35 were classified as background noise, while those with an intensity of 0.35 or higher were extracted as tissue signals. The SNR was defined as $20\log_{10}\frac{\mu}{\sigma}$, where $\mu$ is the mean amplitude of the tissue signals, $\sigma$ is the standard deviation of the background noise.

\section{Results}
\subsection{Digital phantoms}
\begin{figure*}[!ht]
\centerline{\includegraphics[width = 0.55\textwidth]{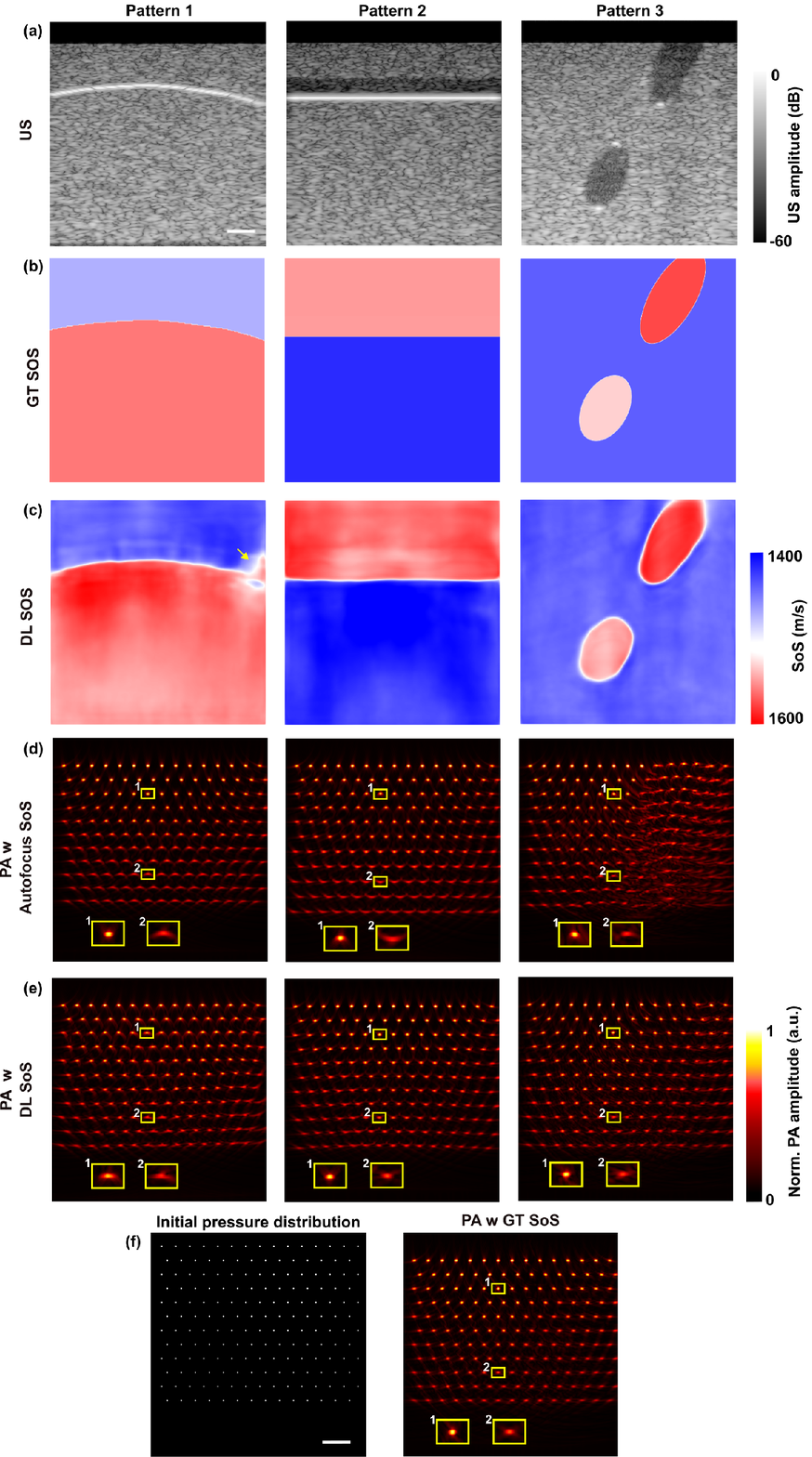}}
\caption{Learning-based sound speed estimation and aberration correction in dual-modal photoacoustic (PA)/ultrasound (US) imaging demonstrated with digital phantoms with point optical absorbers. (a) US reconstruction. The top 2 mm are zeroed out due to the noise interference. (b) The ground truth (GT) of the Sound of Speed (SoS) and (c) SoS distributions predicted by the deep learning (DL) model. (d-e) PA reconstruction using an optimal SoS value given by an autofocus method and SoS distributions by the DL model. (f) Initial pressure distribution used for acoustic forwarding in the k-Wave simulation and PA reconstruction using the GT SoS. The reconstructed optical point sources at two different depths in PA are enlarged in the insets (d-f). Images share the same scale bar of 5 mm.}
\label{fig2-digitalphantom}
\end{figure*}

\begin{table*}[h]
\caption{Root Mean Square Error (RMSE) and Structure Similarity Index Measure (SSIM) Comparison of SoS-Compensated Photoacoustic Images of Digital Phantoms using Autofocus and Deep Learning (DL) Methods (Best global performance was highlighted)}
\centering
\scalebox{0.9}{\begin{tabular}{llcccl}
\cline{3-6}
\multicolumn{2}{l}{\multirow{2}{*}{}}                                             & \multicolumn{2}{c}{RMSE (m/s)}       & \multicolumn{2}{c}{SSIM}             \\ \cline{3-6} 
\multicolumn{2}{l}{}                                                              & Autofocus      & DL            & Autofocus & DL                       \\ \hline
\multicolumn{1}{l|}{\multirow{3}{*}{Pattern 1}} & \multicolumn{1}{l|}{Layer 1}    & 6.23 ± 1.44    & 38.37 ± 19.58 & 0.94      & \multicolumn{1}{c}{0.78} \\ \cline{2-6} 
\multicolumn{1}{l|}{}                           & \multicolumn{1}{l|}{Layer 2}    & 77.72 ± 1.89   & 17.63 ± 14.72 & 0.66      & \multicolumn{1}{c}{0.78} \\ \cline{2-6} 
\multicolumn{1}{l|}{}                           & \multicolumn{1}{l|}{Global}       & 65.11 ± 38.43  & \textbf{25.68 ± 24.19} & \textbf{0.78}      & 0.77                     \\ \hline
\multicolumn{1}{l|}{\multirow{3}{*}{Pattern 2}} & \multicolumn{1}{l|}{Layer 1}    & 17.80 ± 0.40   & 22.44 ± 12.23 & 0.83      & \multicolumn{1}{c}{0.82} \\ \cline{2-6} 
\multicolumn{1}{l|}{}                           & \multicolumn{1}{l|}{Layer 2}    & 102.86 ± 23.56 & 12.75 ± 11.73 & 0.68      & 0.81                     \\ \cline{2-6} 
\multicolumn{1}{l|}{}                           & \multicolumn{1}{l|}{Global}       & 85.22 ± 58.47  & \textbf{18.14 ± 17.62} & 0.76      & \textbf{0.81}                     \\ \hline
\multicolumn{1}{l|}{\multirow{3}{*}{Pattern 3}} & \multicolumn{1}{l|}{Background} & 11.43 ± 2.50   & 9.18 ± 7.83   & 0.73      & \multicolumn{1}{c}{0.87} \\ \cline{2-6} 
\multicolumn{1}{l|}{}                           & \multicolumn{1}{l|}{Inclusions} & 126.18 ± 26.23 & 15.67 ± 5.49  & 0.54      & 0.82                     \\ \cline{2-6} 
\multicolumn{1}{l|}{}                           & \multicolumn{1}{l|}{Global}       & 45.76 ± 38.25  & \textbf{10.21 ± 9.33}  & 0.69      & \textbf{0.86}                     \\ \hline
\end{tabular}}
\label{tab1-numerical}
\end{table*}

\begin{table}[h]
\caption{Statistical analysis of SoS-Compensated Photoacoustic Images of Digital Phantoms using Deep Learning (DL) Methods (n: number of samples)}
\centering
\begin{tabular}{llll}
\hline
           & \thead{Pattern 1 \\ (N = 40)}& \thead{Pattern 2 \\ (N = 40)}& \thead{Pattern 3 \\ (N = 40)}\\ \hline
RMSE (m/s) & 27.95 ± 8.61  & 32.25 ± 7.83 & 12.18 ± 3.74\\ \hline
\end{tabular}
\label{tab2-numerical}
\end{table}

Fig. \ref{fig2-digitalphantom} demonstrated three representative results from digital phantoms. Pattern 1 featured a 2-layer structure with a curved boundary and an uniform echogenicity distribution. The DL model was able to detect SoS discrepancy between two layers, disregarding their identical speckle contrast, with RMSE of 38.37 m/s and 17.63 m/s for each respective layer. In contrast, the global SoS optimisation regarding the image sharpness of PA reconstruction by the autofocus method \cite{treeby2011automatic} resulted in RMSE of 6.23 m/s and 77.72 m/s for each layer, respectively. The implementation details of the autofocus method can be found in Supplementary Materials, Sec.2. On a global scale, the DL method achieved a smaller RMSE of 25.68 m/s compared to 65.11 m/s obtained by the autofocus method. In addition, it demonstrated a comparable SSIM of 0.77, as opposed to 0.78 achieved by the autofocus method. The SoS-compensated PA reconstruction was shown in Fig. \ref{fig2-digitalphantom} (d) and (e), first column. As expected, the performance of aberration correction with the DL method was degraded at larger depths (lager than 3 cm). This could be attributed to the SoS estimation errors near the tissue boundary (denoted by the yellow arrows), which was probably due to the presence of the reflection artefacts. 

Pattern 2 depicted the US anatomy with straight surfaces and isoechoic layers. Despite the echogenicity varying in the top two layers, the SoS prediction remained consistent using the DL method, with RMSE of 22.44 m/s. For the bottom layer (Layer 2, as indicated in Tab. \ref{tab1-numerical}), the DL method yielded a much smaller RMSE of 12.75 m/s compared to 102.86 m/s obtained by the autofocus method. Thus, the PA reconstruction was improved by incorporating the DL SoS, as shown in Fig. \ref{fig2-digitalphantom} (e), middle column, resulting in an SSIM of 0.81 compared to 0.76 obtained by the autofocus method. Besides, Pattern 3 mimics a US anatomy, featuring a homogenous background and hypoechoic elliptical inclusions. The DL model effectively retrieved the SoS distributions of both the background and the inclusions, achieving RMSEs as small as 9.18 m/s and 15.67 m/s for the background and inclusions, respectively. The enhancement in PA image reconstruction was also observed as shown in Fig. \ref{fig2-digitalphantom} (e), right column. However, the autofocus method introduced large errors (up to 126.18 m/s in RMSE) for the inclusions, as evidenced by the serve distortions in the PA image reconstruction (Fig. \ref{fig2-digitalphantom} (d), right column). Furthermore, the generalisation performance of the DL model was statistically analysed using a total of 120 samples (40 samples randomly selected from each representative pattern) and summarised in Tab. \ref{tab2-numerical}. The achieved RMSEs were close to those acquired during the model training, with the smallest RMSE of 12.18 m/s observed in the group with US anatomy assembled to Pattern 3. 

\subsection{Agar-based tissue-mimicking phantoms}
\begin{figure*}[!ht]
\centerline{\includegraphics[width= \textwidth]{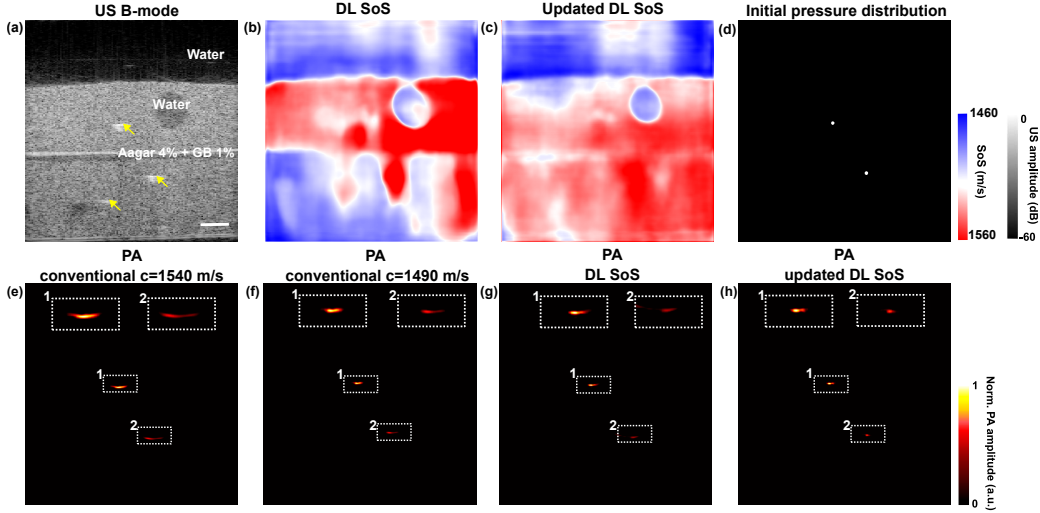}}
\caption{Learning-based sound speed estimation and aberration correction in dual-modal photoacoustic (PA)/ultrasound (US) imaging demonstrated with agar-based tissue-mimicking phantom with pencil leads as inclusions (indicated by yellow arrows in US B-mode reconstruction). (a) US B-mode image. (b-c) SoS distributions predicted by the deep learning (DL) model. (d) Initial pressure distribution. (e-h) PA reconstructions using the SoS of soft tissues (c = 1540 m/s) and water (c = 1490 m/s), and the SoS distributions by the DL model. The reconstructed pencil leads (cross-sections) in PA are enlarged in the insets. Scale bar: 5 mm.}
\label{fig2-agarphantom}
\end{figure*}

\begin{table*}[h]
\caption{Statistical analysis of DL-based SoS prediction using agar-based tissue-mimicking phantoms (N: number of samples. N = 60)}
\centering
\begin{tabular}{c|ll}
\hline
\multicolumn{1}{l|}{RMSE (m/s)} & \multicolumn{1}{c}{Base model} & \multicolumn{1}{c}{Base model (after TL)} \\ \hline
Water                           & 44.21 ± 22.07        & \textbf{15.48 ± 1.49}       \\ \hline
Agar 2\%                        & 40.39 ± 19.33        & \textbf{26.75 ± 7.68}       \\ \hline
Agar 4\%                        & 41.89 ± 20.41        & \textbf{15.98 ± 1.11}       \\ \hline
Agar 6\%                        & 37.67 ± 16.73        & \textbf{15.24 ± 5.67}       \\ \hline
\end{tabular}
\label{tab-phantom}
\end{table*}

An exemplar result is shown in Fig. \ref{fig2-agarphantom}. As shown in the B-mode image, the phantom consisted of two layers made with the same agar concentrations of 4\%. A tubular structure with a diameter of around 5 mm was introduced at the upper layer. Meanwhile, three pencil leads serving as the optical absorbers appeared in the B-mode image with a higher speckle intensity than the background, as indicated by the yellow arrows.  

In Fig. \ref{fig2-agarphantom}, the aberration artefacts were observed in the PA reconstruction with the conventional SoS assumption of 1540 m/s. The cross-sections of the pencil leads were distorted. For point source 1, the lateral FWHM was 1.52 mm compared to the actual dimension of 0.50 mm (lateral resolution of the imaging system: 0.59 mm \cite{xia2018handheld}). For point source 2 at a deeper depth, the aberration was accumulated, resulting in a lateral FWHM of 2.47 mm. With a uniform SoS of 1490 m/s, the aberration caused by the coupling medium, i.e., water was compensated. The lateral FWHM for the point source 1 and the point source 2 was reduced to 1.15 mm and 1.60 mm, respectively. The PA signals were then reconstructed with the SoS parameterised using the DL output. With the SoS map predicted by the base model (without transfer training using phantom data), the lateral FWHMs of points 1 and 2 were 1.13 mm and 1.29 mm, respectively. However, the performance was degraded at deep depths, especially for the bottom layer, with an average SoS of 1477.5 m/s compared to the measured 1520.3 m/s. In contrast, the updated model demonstrated good performance of SoS prediction at deep depths. The average SoS values for the coupling layer, upper layer, and bottom layer were 1484.4 m/s, 1521.8 m/s, and 1528 m/s, respectively. The lateral resolution, represented by the lateral FWHMs, was improved to 0.97 mm and 0.74 mm for points 1 and 2, respectively. Furthermore, local SSIM based on the region of interests (ROIs) that enclosed the point sources was calculated using the initial pressure distribution of the phantom for reference, as shown in Fig. \ref{fig2-agarphantom}(d). The point sources were segmented from the US B-mode image, with a physical dimension of 0.5 mm and an initial pressure of 1. The SSIM values were 0.78 and 0.84 for the PA images reconstructed using the conventional SoS assumption of 1540 m/s and the SoS of water 1490 m/s, respectively. The highest SSIM of 0.88 was achieved using the SoS distribution estimated by the DL model after transfer training.

The model was evaluated using different agar-based tissue-mimicking phantoms. Additional results were included in Supplementary Materials, Fig. s4. Tab. \ref{tab-phantom} compared the estimation errors of SoS for different media including the coupling water and tissue-mimicking phantoms containing different agar concentrations. A total of 60 measurements (20 measurements from the same phantom) were used for statistical analysis. 

\begin{figure*}[!ht]
\centerline{\includegraphics[width= \textwidth]{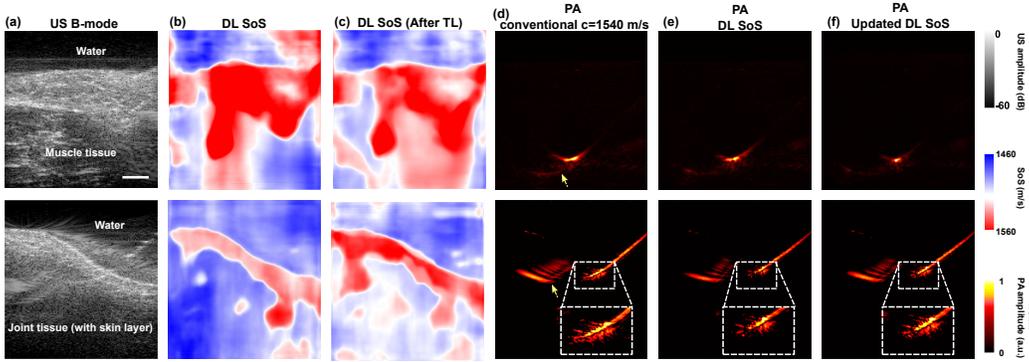}}
\caption{Learning-based sound speed esitmation and aberration correction in dual-modal photoacoustic (PA)/ultrasound (US) imaging demonstrated with needle in-plane and out-of-plane insertions into porcine tissues ex vivo. (a) US B-mode image. (b-c) SoS distributions predicted by the deep learning (DL) model. (d-e) PA reconstructions using a conventional SoS value in soft tissue (c = 1540 m/s) and the SoS distributions by the DL model; Yellow arrows denote image artefacts associated with acoustic reflection and limited view. Scale bar: 5 mm.}
\label{fig3-exvivoporkbelly}
\end{figure*}

\subsection{Ex vivo tissue experiments simulating interventional procedures}
Acoustic aberration can distort the PA visualisation of medical devices, such as clinical metallic needles that are commonly used during PA/US-guided minimally invasive procedures \cite{xia2018handheld}. In Fig. \ref{fig3-exvivoporkbelly} (d), a clinical spinal needle with enhanced tip visualisation \cite{shi2022enhanced} was inserted into the muscle tissue using a out-of-plane approach (upper). The tip signals in the PA reconstructions were distorted with a uniform SoS of 1540 m/s; for the in-plane needle insertion (bottom), the distal segment of the needle experienced a bending distortion (shown in the zoom-in images). Both of them led to ambiguity in tip localisation under PA guidance. The estimated SoS of water was 1497.7 m/s, matching the literature value at 20 $^\circ C$ (1490 m/s). Besides, the SoS for muscle tissue and the compound layer of skin and fat in the joint tissue were estimated as 1547.6 m/s and 1558.6 m/s, respectively. The estimations were acceptable, given the known variation in SoS, ranging from 1436 m/s to 1470 m/s for the fat layers and 1682 m/s for skin \cite{koch2011ultrasounda}. By employing the US-informed SoS compensation for PA reconstruction, the aberration artefacts at the tip area were suppressed, as shown in Fig. \ref{fig3-exvivoporkbelly} (e-f). Furthermore, compared to the base model trained only from the simulated data (SoS distributions were shown in Fig. \ref{fig3-exvivoporkbelly}(b)), the transfer learning model demonstrated improved performance on SoS predictions (Fig. \ref{fig3-exvivoporkbelly}(c)).

\subsection{In vivo human data}
\begin{figure*}[!ht]
\centerline{\includegraphics[width= \textwidth]{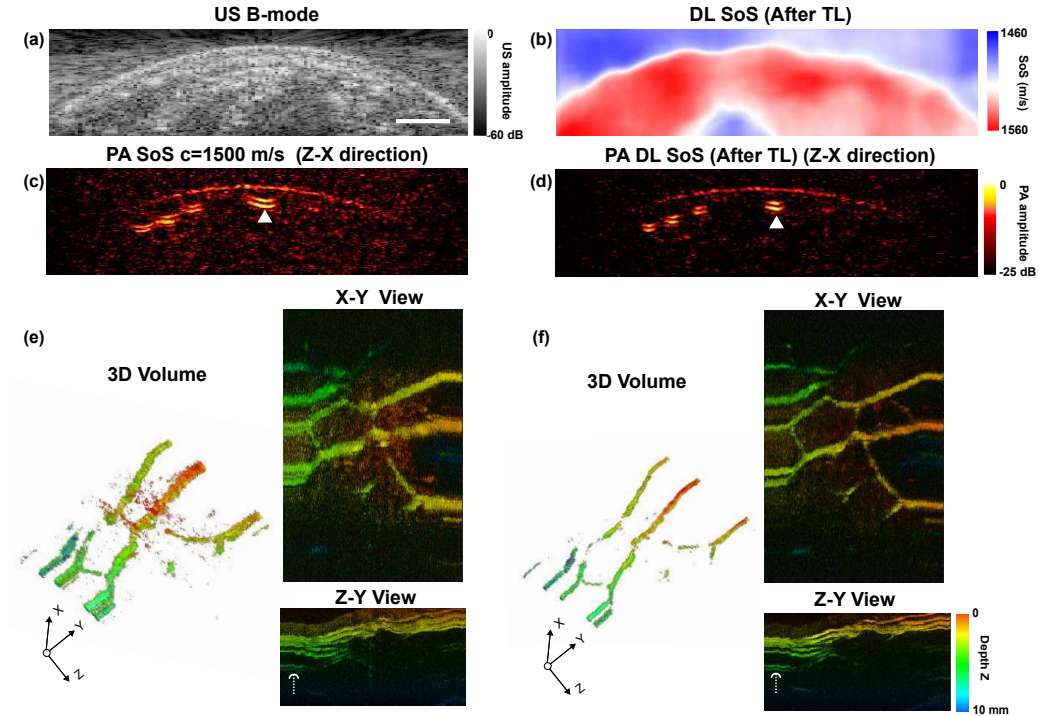}}
\caption{Learning-based sound speed estimation and aberration correction in dual-modal photoacoustic (PA)/ultrasound (US) imaging demonstrated with in vivo 3D volume of the human wrist. (a) US B-mode image. (b) SoS distribution by the deep learning (DL) model after transfer training (c-d) PA reconstructions using a uniform SoS value of water (c = 1500 m/s at 23 $^\circ$C) and the SoS distribution by the DL model. (e-f) 3D volumes and maximum intensity projections from X-Y and Z-Y directions.
White triangular denotes the cross-section of a blood vessel. White arrow denotes vessels at deep depths. Scale bar: 5 mm.}
\label{fig4-invivo}
\end{figure*}

A 3D scanning sequence consisting of 384 2D imaging slices of a human wrist was used to evaluate the model's performance, focusing on its reproducibility on the data of low SNRs. The recovered SoS value of the muscle tissue by the DL framework was 1546.0 m/s $\pm$ 6.9 m/s, well corresponding to the known SoS of human soft tissues (~1540 m/s) \cite{mamou2013quantitative}. The water SoS was 1502.7 m/s $\pm$ 4.2 m/s. In the PA reconstructions, the cross-section signals of blood vessels were distorted laterally (indicated by white triangulars in Fig. \ref{fig4-invivo} (c-d)) with a uniform SoS assumption of 1500 m/s, which was determined based on the measured water temperature, i.e. 23 $^\circ C$). In contrast, the PA images compensated by the DL SoS enhanced the visualisation of the blood vessels by suppressing the aberration artefacts. Herein, the SNR was improved from 18.4 dB $\pm$ 0.98 to 20.4 dB $\pm$ 0.58 (p$<$0.001). 

The enhancement in the superficial vasculature reconstruction using PA imaging can be observed with the depth-coded 3D reconstructions and the corresponding maximum intensity projections from X-Y and Z-Y direction, respectively (shown in Fig. \ref{fig4-invivo} (e-f)). The DL-based SoS compensation can be potentially helpful for visualising small vessels where PA signals from the vessels can be readily contaminated by noise, or vessels at deep depths where the signals were diminished due to significant light attenuation (denoted by white arrows in Fig. \ref{fig4-invivo} (e-f) Z-Y view). 

\section{Discussion}
A long-standing challenge in PA image reconstruction is the lack of accurate information of SoS variations in heterogeneous biological tissue. Failure to compensate for the variation in SoS can result in severe image distortions and artefacts, which compromise the image quality. Prior studies have explored different approaches to incorporate the SoS heterogeneity during PA image reconstruction. Although promising results have been reported, they usually involve certain complexities such as dedicated imaging hardware and sophisticated algorithm developments. Recent works demonstrated that deep learning was able to directly retrieve SoS distributions from raw channel data of pulse-echo US imaging \cite{feigin2020deep, vishnevskiy2018image,jush2020dnnbased}. Feign et al. proposed an SoS inversion method using a fully convolutional deep neural network \cite{feigin2020deep}. The network was trained using simulated plane wave US raw data. Results on the human neck and calf muscles provided promising indications of SoS variations. Similarly, Jush et al. explored a deep neural network for SoS reconstruction for US breast imaging using single plane wave US acquisition \cite{jush2020dnnbased} and further extended to in-phase and quadrature data as the input \cite{khunjush2021datadriven}. Inspired by the feasibility of retrieving SoS distributions from US channel data with deep neural networks, a learning-based SoS correction method for PA imaging was proposed based on a dual-modal PA/US imaging system. 

The dual-modal system can acquire interleaved PA and US data for real-time applications, making it possible to enhance PA image reconstruction using SoS information inherited in the corresponding co-registered US data. The dataset for training a fully convolutional deep neural network was prepared in silico in 2D, taking into consideration the variations of US anatomy and SoS values as well as computational efficiency. Noise including thermal noise and system noise from real measurements was added. Even though the digital phantoms used for model evaluation contain anatomical structures and echogenicity patterns that differ from the training distribution, the trained network can estimate their SoS distributions with high fidelity. The estimated SoS maps can compensate for whole-field SoS variations during PA reconstruction. In comparison, conventional and an autofocus method \cite{treeby2011automatic} based on maximising image sharpness resulted in deformations in point source reconstructions as evidenced by degradation in the lateral resolution. 

The trained network had a degraded performance when applied to measurement data, which could be attributed to its poor robustness against out-of-plane artefacts and system noise. It was observed that transfer learning using even a small amount of data acquired from physical phantoms was able to improve the model performance. The estimated SoS values for soft tissues were situated in the range of the values reported in the literature and were found to be effective in mitigating the SoS aberration artefacts in the conventional PA reconstruction. In particular, the network can identify the SoS inconsistency between coupling medium (water) and soft tissues, which was of importance to correct the positions of medical devices such as metallic needles. This could be advantageous for tracking the needle tip relative to the patient anatomy during various US-guided minimally invasive procedures such as peripheral nerve blocks and tumour biopsy \cite{shi2022enhanced, xia2015inplane, xia2015interventional,baker2022intraoperative}. Also, the network after transfer training generalised well when applied to in vivo human data. The SNR of the imaged vasculature improved from 18.4 dB to 20.4 dB using the proposed framework. This is particularly useful for LED-based systems where small capillaries can be readily suppressed by thermal noise due to the low optical fluence. 

The network took around 0.01s for a single frame SoS estimation with a GPU (Tesla T4), indicating the potential for real time PA image reconstruction with SoS compensation. The current implementation of TR-based PA image reconstruction in k-Wave required approximately 1 mins per single frame with sizes of 788 $\times$ 768 with an NVIDIA Quadro RTX 5000 GPU. Future works could include the development of a fast image reconstruction algorithm that incorporates SoS distributions in tissue such as a deep neural network \cite{hauptmann2018modelbased, dehner2023deep}. 
 
The accuracy in terms of SoS estimation by the networks can be improved from several perspectives. For example, a single wave transmission may not be adequate for a full-field recovery, which is also reported in \cite{feigin2020deep}. Advanced US transmission strategies such as coherent compounding could be potentially helpful for acquiring data with improved SNRs \cite{entrekin2001realtime}. The coherency across channels can also be explored. In addition, inhomogeneous acoustic attenuation could be incorporated during US simulations. 

\section{Conclusions}

 We developed a deep-learning framework for accurate PA image reconstruction with SoS compensation on a dual-modal PA/US imaging system exploiting a clinical US probe, which is particularly suitable for real-time applications. It can estimate SoS distributions using US RF channel data and correct aberration artefacts in PA images caused by SoS heterogeneity. With transfer learning using tissue-mimicking phantoms, the deep learning model exhibited improved robustness to experimental ex vivo and in vivo data. Furthermore, image enhancement in PA reconstructions was observed in a human volunteer study compared to conventional reconstructions. Thus, this framework holds potential for diverse preclinical and clinical applications, including the facilitation of minimally invasive medical procedures.

\section{Declaration of competing interests}
The authors declare that they have no known competing financial interests or personal relationships that could have appeared to influence the work reported in this paper. T.V. is co-founder and shareholder of Hypervision Surgical Ltd, London, UK. 

\section{Acknowledgements}
This project was supported by the Wellcome Trust, United Kingdom (203148/Z/16/Z), the Engineering and Physical Sciences Research Council (EPSRC), United Kingdom (NS/A000049/1), and King’s–China Scholarship Council PhD Scholarship Program (K-CSC) (202008060071). For the purpose of open access, the authors have applied a CC BY public copyright license to any author-accepted manuscript version arising from this submission.




\bibliographystyle{elsarticle-num} 
\bibliography{Sound-speed-correction-v1,additions-background}

\end{document}